\documentclass[prd,reprint,showpacs,showkeys]{revtex4}
\usepackage{amsfonts}
\usepackage{amssymb}
\usepackage{amsmath}
\usepackage{graphicx}
\usepackage[font={footnotesize,it}]{caption}

\begin{document}

\title{Generalized Monge gauge}
\author{S. Habib Mazharimousavi}
\email{habib.mazhari@emu.edu.tr}
\author{S. Danial Forghani}
\email{danial.forghani@emu.edu.tr}
\author{S. Niloufar Abtahi}
\email{sayedeh.abtahi@emu.edu.tr}
\affiliation{Department of Physics, Eastern Mediterranean University, G. Magusa, north
Cyprus, Mersin 10, Turkey. }
\date{\today }

\begin{abstract}
Monge gauge in differential geometry is generalized. The original Monge
gauge is based on a surface defined as a height function $h(x,y)$ above a
flat reference plane. The total curvature and the Gaussian curvature are
found in terms of the height function. Getting benefits from our
mathematical knowledge of general relativity, we shall extend the Monge
gauge toward more complicated surfaces. Here in this study we consider the
height function above a curved surface namely a sphere of radius $R$. The
proposed height function is a function of $\theta $ and $\varphi $ on a
closed interval. We find the first, second fundamental forms and the total
and Gaussian curvatures in terms of the new height function. Some specific
limits are discussed and two illustrative examples are given.
\end{abstract}

\maketitle

\section{Introduction}

Applications of pure mathematics in applied science are not rare. For
decades differential geometry has been known as a useful tool to study lipid
membranes \cite{0}. Tu and Ou-Yang \cite{Tu} have published a review work on
the recent theoretical advances in elasticity of membranes\ which covers
most of the recent and initial papers in this context. Another review paper
on the same issue has been published by Deserno in \cite{1}. In this work,
by using differential geometry, fluid lipid membrane as a $2-$dimensional
surface has been investigated. In \cite{G} the differential geometry of
proteins and its applications have been presented. Application of helicoidal
membrane in connecting the stacked endoplasmic reticulum sheets inside the
cell has been worked out by Terasaki et al in \cite{T} and highlighted by
Marshall in \cite{Cell}. To the first order approximation, biological
membranes are made of a bilayer of phospholipid molecules embedded in water 
\cite{1,2,3}. The physics of cell membranes including physics behind the
self-assembly, molecular structures and electrical properties are given in 
\cite{PCM}.

Far from the biological (cell-) membranes, in the Einstein theory of gravity
there are objects which are called timelike thin-shells. These are basically
membranes in spacetime. As the theory of general relativity itself, the
mathematical structures of such objects are very rich. Although, initially
the spacetime has been assigned to be $3+1-$dimensional but the lower
dimensional spacetime i.e., $2+1$ is also very well known model. The analogy
between the $1+1-$dimensional thin-shells in $2+1-$dimensional spacetime and
the $2-$dimensional surfaces in $3-$dimensional space is the point that we
shall get into in our formalism. More precisely, in this paper, we shall try
to borrow concepts / mathematical rules of general relativity mainly in
connections to the concept of thin-shells to develop the basic tools used in
the study of (cell-) membranes. One of the most used concept in the early
development of the biological membranes is the so called Monge Gauge (MG).
The MG \cite{1} is a parametrization / mapping of a $2-$dimensional surface
which is defined by a height function $h\left( x,y\right) $ over a flat
plane as a function of orthonormal coordinate on the plane $x$ and $y$ into
a $3-$dimensional flat space of coordinates $x,y$ and $z.$ This is usually
shown as%
\begin{equation}
h:\left\{ 
\begin{array}{c}
\mathbb{R}
^{2}\supset U\longrightarrow 
\mathbb{R}
^{3} \\ 
\left( x,y\right) \longrightarrow h\left( x,y\right)%
\end{array}%
\right. .
\end{equation}%
Using the standard definition of the total and the Gaussian curvature one
finds \cite{1} 
\begin{equation}
\kappa =\frac{h_{,xx}\left( 1+h_{,y}^{2}\right) +h_{,yy}\left(
1+h_{,x}^{2}\right) -2h_{,xy}h_{,x}h_{y}}{2\left(
1+h_{,x}^{2}+h_{,y}^{2}\right) ^{2}}
\end{equation}%
and%
\begin{equation}
\kappa _{G}=\frac{h_{,xx}h_{,yy}-h_{,xy}^{2}}{\left(
1+h_{,x}^{2}+h_{,y}^{2}\right) ^{2}}
\end{equation}%
respectively. Introducing $\mathbf{\nabla }=\left( 
\begin{array}{c}
\frac{\partial }{\partial x} \\ 
\frac{\partial }{\partial y}%
\end{array}%
\right) $ the above usually are compacted as%
\begin{equation}
2\kappa =\mathbf{\nabla }.\left( \frac{\mathbf{\nabla }h}{\sqrt{1+\left( 
\mathbf{\nabla }h\right) ^{2}}}\right)
\end{equation}%
and%
\begin{equation}
\kappa _{G}=\frac{\det \left( \partial ^{2}h\right) }{\left( 1+\left( 
\mathbf{\nabla }h\right) ^{2}\right) ^{2}}
\end{equation}%
in which the Hessian $\partial ^{2}h$ is given by%
\begin{equation}
\partial ^{2}h=\left( 
\begin{array}{cc}
h_{,xx} & h_{,xy} \\ 
h_{,yx} & h_{,yy}%
\end{array}%
\right) .
\end{equation}%
For instance, in the famous paper of Seifert and Langer \cite{4}, the height
function was set to $h\left( x,y\right) =h\exp \left[ iqx\right] +c.c.$ in
which $h$ and $q$ are two constants. For more recent work one may look at
the work of Bingham, Smye and Olmsted \cite{5}.

As of \cite{4} and \cite{5} the other studies where the MG has been used,
the original unperturbed surface is locally flat. However, in the case of a
blood cell we may not be able to consider its membrane flat. Therefore and
fluctuation from its original rest shape needs to be analyzed exactly
without using MG. In this study we show that in such cases one may use a
proper MG which of course is not the one introduced above. Spherically
symmetry is the most common symmetry which occurs in nature and hence in
this study we concentrate on spherically symmetric MG.

\section{The spherical MG}

Let's start with a spherical shell of radius $R$ with some fluctuation on
its surface given by the spherical height function $h\left( \theta ,\varphi
\right) $ in which $\theta $ and $\varphi $ are the polar and azimuthal
angle. Unlike the Cartesian height function, the non fluctuated sphere is
given by $h\left( \theta ,\varphi \right) =R.$ Our aim is to find the first
and second fundamental forms as well as the extrinsic curvature tensor and
the scalar curvature of the hypersurface $\Sigma $ in terms of $h\left(
\theta ,\varphi \right) $ only. We start with the definition of the normal
vector on the surface $\Sigma $ given by%
\begin{equation}
n_{\gamma }=\left. \frac{1}{\sqrt{\Delta }}\frac{dF}{dx^{\gamma }}%
\right\vert _{\Sigma }
\end{equation}%
in which $F:=r-h\left( \theta ,\varphi \right) =0$ is the definition of the
surface $\Sigma $ in three dimensional flat spherical symmetric bulk space $%
M $ with line element%
\begin{equation}
ds_{M}^{2}=g_{\alpha \beta }dx^{\alpha }dx^{\beta }=dr^{2}+r^{2}\left(
d\theta ^{2}+\sin ^{2}\theta d\varphi ^{2}\right) .
\end{equation}%
Herein $\Delta $ is defined as%
\begin{equation}
\Delta =\left. g^{\alpha \beta }\frac{dF}{dx^{\alpha }}\frac{dF}{dx^{\beta }}%
\right\vert _{\Sigma }
\end{equation}%
in which $g^{\alpha \beta }$ is the metric tensor of the bulk and $n_{\gamma
}n^{\gamma }=1$. Let's note that the Greek letters $\alpha ,\beta ,...=1,2,3$
are used for the bulk space while the Latin letters $i,j,...=2,3$ shall be
used for the hypersurface. Using the definition (1) we find%
\begin{equation}
n_{r}=\frac{1}{\sqrt{\Delta }}
\end{equation}%
\begin{equation}
n_{\theta }=\frac{-1}{\sqrt{\Delta }}h_{,\theta }
\end{equation}%
and%
\begin{equation}
n_{\varphi }=\frac{-1}{\sqrt{\Delta }}h_{,\varphi }
\end{equation}%
with%
\begin{equation}
\Delta =1+\frac{h_{,\theta }^{2}}{h^{2}}+\frac{h_{,\varphi }^{2}}{h^{2}\sin
^{2}\theta }.
\end{equation}%
The induced metric on the hypersurface $\Sigma $ is given by%
\begin{equation}
g_{ij}=\frac{\partial x^{\alpha }}{\partial \xi ^{i}}\frac{\partial x^{\beta
}}{\partial \xi ^{j}}g_{\alpha \beta }
\end{equation}%
in which the line element on the hypersurface is written as%
\begin{equation}
ds_{\Sigma }^{2}=g_{ij}d\xi ^{i}d\xi ^{j}.
\end{equation}%
Explicitly one finds%
\begin{equation}
ds_{\Sigma }^{2}=\left( h^{2}+h_{,\theta }^{2}\right) d\theta ^{2}+\left(
h^{2}\sin ^{2}\theta +h_{,\varphi }^{2}\right) d\varphi ^{2}+2h_{,\theta
}h_{,\varphi }d\theta d\varphi
\end{equation}%
or simply%
\begin{equation}
g_{ij}=\left[ 
\begin{array}{cc}
h^{2}+h_{,\theta }^{2} & h_{,\theta }h_{,\varphi } \\ 
h_{,\theta }h_{,\varphi } & h^{2}\sin ^{2}\theta +h_{,\varphi }^{2}%
\end{array}%
\right]
\end{equation}%
with its inverse%
\begin{equation}
g^{ij}=\frac{1}{h^{4}\sin ^{2}\theta \Delta }\left[ 
\begin{array}{cc}
h^{2}\sin ^{2}\theta +h_{,\varphi }^{2} & -h_{,\theta }h_{,\varphi } \\ 
-h_{,\theta }h_{,\varphi } & h^{2}+h_{,\theta }^{2}%
\end{array}%
\right] .
\end{equation}%
Having the induced metric and the normal vector on the hypersurface $\Sigma
, $ one can use the definition of extrinsic curvature tensor of the surface $%
\Sigma $ 
\begin{equation}
K_{ij}=\left. -n_{\gamma }\left( \frac{\partial ^{2}x^{\gamma }}{\partial
\xi ^{i}\partial \xi ^{j}}+\Gamma _{\alpha \beta }^{\gamma }\frac{\partial
x^{\alpha }}{\partial \xi ^{i}}\frac{\partial x^{\beta }}{\partial \xi ^{j}}%
\right) \right\vert _{\Sigma }
\end{equation}%
to find the nonzero components of the extrinsic curvature or second
fundamental form. We note that $\Gamma _{\alpha \beta }^{\gamma }$ are the
Christoffel symbols of the second kind of the bulk space with the only
nonzero components given by%
\begin{equation}
\Gamma _{r\theta }^{\theta }=\Gamma _{\theta r}^{\theta }=\Gamma _{r\varphi
}^{\varphi }=\Gamma _{\varphi r}^{\varphi }=\frac{1}{r}
\end{equation}%
\begin{equation}
\Gamma _{\theta \theta }^{r}=-r,\text{ }\Gamma _{\varphi \varphi
}^{r}=-r\sin ^{2}\theta ,\text{ }
\end{equation}%
and%
\begin{equation}
\Gamma _{\varphi \theta }^{\varphi }=\Gamma _{\theta \varphi }^{\varphi }=%
\frac{\cos \theta }{\sin \theta },\text{ }\Gamma _{\varphi \varphi }^{\theta
}=-\sin \theta \cos \theta .
\end{equation}%
The components of the extrinsic curvature tensor are found to be%
\begin{equation}
K_{\theta \theta }=\frac{h^{2}+2h_{,\theta }^{2}-hh_{,\theta \theta }}{h%
\sqrt{\Delta }}
\end{equation}%
\begin{equation}
K_{\varphi \varphi }=\frac{h^{2}\sin ^{2}\theta -hh_{,\theta }\sin \theta
\cos \theta +2h_{,\varphi }^{2}-hh_{,\varphi \varphi }}{h\sqrt{\Delta }}
\end{equation}%
\begin{equation}
K_{\varphi \theta }=K_{\theta \varphi }=\frac{2h_{,\theta }h_{,\varphi
}+hh_{,\varphi }\frac{\cos \theta }{\sin \theta }-hh_{,\theta \varphi }}{h%
\sqrt{\Delta }}.
\end{equation}%
Next, we find $K_{i}^{j}$ in order to find the total and the Gaussian
curvature. Using the induced metric one finds%
\begin{equation}
K_{\theta }^{\theta }=\frac{\left( h^{2}-hh_{,\theta \theta }+2h_{,\theta
}^{2}\right) h\sin ^{3}\theta +\left( hh_{,\varphi }^{2}+h_{,\theta
}h_{,\varphi }h_{,\theta \varphi }-h_{,\varphi }^{2}h_{,\theta \theta
}\right) \sin \theta -h_{,\theta }h_{,\varphi }^{2}\cos \theta }{h^{4}\sin
^{3}\theta \Delta ^{3/2}},
\end{equation}%
\begin{equation}
K_{\varphi }^{\varphi }=\frac{\left( h^{2}+h_{,\theta }^{2}\right) \left(
h\sin ^{2}\theta -h_{,\theta }\sin \theta \cos \theta -h_{,\varphi \varphi
}\right) \sin \theta +\left( 2hh_{,\varphi }^{2}+h_{,\theta }h_{,\varphi
}h_{,\theta \varphi }\right) \sin \theta -h_{,\theta }h_{,\varphi }^{2}\cos
\theta }{h^{4}\sin ^{3}\theta \Delta ^{3/2}},
\end{equation}%
\begin{equation}
K_{\varphi }^{\theta }=\frac{\left( h_{,\theta }h_{,\varphi }-hh_{,\theta
\varphi }\right) h\sin ^{3}\theta +h_{,\varphi }\left( h_{,\theta
}^{2}+h^{2}\right) \sin ^{2}\theta \cos \theta +\left( h_{,\theta
}h_{,\varphi \varphi }-h_{,\varphi }h_{,\theta \varphi }\right) h_{,\varphi
}\sin \theta +h_{,\varphi }^{3}\cos \theta }{h^{4}\sin ^{3}\theta \Delta
^{3/2}}
\end{equation}%
and%
\begin{equation}
K_{\theta }^{\varphi }=\frac{\left( h_{,\theta }^{2}+h^{2}\right) \left(
-h_{,\theta \varphi }\sin \theta +h_{,\varphi }\cos \theta \right) +\sin
\theta \left( h_{,\theta }h_{,\varphi }\left( h+h_{,\varphi \varphi }\right)
\right) }{h^{4}\sin ^{3}\theta \Delta ^{3/2}}.
\end{equation}%
Finally we find the mean curvature $\kappa =\frac{1}{2}tr\left(
K_{i}^{j}\right) $ and the Gaussian curvature $\kappa _{G}=\det \left(
K_{i}^{j}\right) ,$ given by%
\begin{equation}
\kappa =\frac{\alpha _{1}\sin ^{3}\theta +\alpha _{2}\sin ^{2}\theta \cos
\theta +\alpha _{3}\sin \theta +\alpha _{4}\cos \theta }{2h^{4}\Delta ^{%
\frac{3}{2}}\sin ^{3}\theta }
\end{equation}%
in which%
\begin{equation}
\left( 
\begin{array}{r}
\alpha _{1} \\ 
\alpha _{2} \\ 
\alpha _{3} \\ 
\alpha _{4}%
\end{array}%
\right) =\left( 
\begin{array}{c}
3hh_{,\theta }^{2}+2h^{3}-h^{2}h_{,\theta \theta } \\ 
-h_{,\theta }\left( h^{2}+h_{,\theta }^{2}\right) \\ 
h_{,\varphi }^{2}\left( 3h-h_{,\theta \theta }\right) +2h_{,\theta
}h_{,\varphi }h_{,\theta \varphi }-h_{,\varphi \varphi }\left(
h^{2}+h_{,\theta }^{2}\right) \\ 
-2h_{,\theta }h_{,\varphi }^{2}%
\end{array}%
\right) ,
\end{equation}%
and%
\begin{equation}
\kappa _{G}=\frac{\left( h_{,\varphi }^{2}+\left( h^{2}+h_{,\theta
}^{2}\right) \sin ^{2}\theta \right) \left( \beta _{1}\sin ^{4}\theta +\beta
_{2}\sin ^{3}\theta \cos \theta +\beta _{3}\sin ^{2}\theta +\beta _{4}\sin
\theta \cos \theta +\beta _{5}\cos ^{2}\theta \right) }{h^{7}\Delta ^{3}\sin
^{6}\theta }
\end{equation}%
in which%
\begin{equation}
\left( 
\begin{array}{r}
\beta _{1} \\ 
\beta _{2} \\ 
\beta _{3} \\ 
\beta _{4} \\ 
\beta _{5}%
\end{array}%
\right) =\left( 
\begin{array}{c}
h\left( h^{2}+2h_{,\theta }^{2}-hh_{,\theta \theta }\right) \\ 
-h_{\theta }\left( h^{2}+2h_{,\theta }^{2}-hh_{,\theta \theta }\right) \\ 
-h_{,\varphi \varphi }h^{2}+\left( h_{,\theta \theta }h_{,\varphi \varphi
}-h_{,\theta \varphi }^{2}+2h_{,\varphi }^{2}\right) h+4h_{,\theta
}h_{,\varphi }h_{,\theta \varphi }-2h_{,\theta }^{2}h_{,\varphi \varphi
}-2h_{,\varphi }^{2}h_{,\theta \theta } \\ 
2h_{,\varphi }\left( -2h_{,\theta }h_{,\varphi }+h_{,\theta \varphi }h\right)
\\ 
-hh_{,\varphi }^{2}%
\end{array}%
\right)
\end{equation}%
respectively.

\section{Special cases}

Our first special case is the perfect sphere with $h\left( \theta ,\varphi
\right) =R.$ This in turn implies $h_{,\theta }=h_{,\varphi }=0$ and
therefore%
\begin{equation}
K_{i}^{j}=\left[ 
\begin{array}{cc}
\frac{1}{R} & 0 \\ 
0 & \frac{1}{R}%
\end{array}%
\right]
\end{equation}%
and therefore $\kappa =\frac{1}{R}$ and $\kappa _{G}=\frac{1}{R^{2}}.$

In the second special case we set $h\left( \theta ,\varphi \right) =h\left(
\varphi \right) $ and consequently $h_{,\theta }=0.$ These result in%
\begin{equation}
g_{ij}=\left[ 
\begin{array}{cc}
h^{2} & 0 \\ 
0 & h^{2}\sin ^{2}\theta +h_{,\varphi }^{2}%
\end{array}%
\right]
\end{equation}%
and%
\begin{equation}
K_{i}^{j}=\left[ 
\begin{array}{cc}
\frac{h^{2}\sin ^{2}\theta +h_{,\varphi }^{2}}{h^{3}\sin ^{2}\theta \Delta
^{3/2}} & \frac{h_{,\varphi }h^{2}\sin ^{2}\theta \cos \theta +h_{,\varphi
}^{3}\cos \theta }{h^{4}\sin ^{3}\theta \Delta ^{3/2}} \\ 
\frac{h_{,\varphi }\cos \theta }{h^{2}\sin ^{3}\theta \Delta ^{3/2}} & \frac{%
h^{2}\sin ^{2}\theta -hh_{,\varphi \varphi }+2h_{,\varphi }^{2}}{h^{3}\sin
^{2}\theta \Delta ^{3/2}}%
\end{array}%
\right]
\end{equation}%
in which $\Delta =1+\frac{h_{,\varphi }^{2}}{h^{2}\sin ^{2}\theta }.$
Following $K_{i}^{j}$ one finds%
\begin{equation}
\kappa =\frac{2h^{2}\sin ^{2}\theta +3h_{,\varphi }^{2}-hh_{,\varphi \varphi
}}{2h^{3}\Delta ^{3/2}\sin ^{2}\theta }
\end{equation}%
and%
\begin{equation}
\kappa _{G}=\frac{h^{2}\sin ^{4}\theta +\left( 2h_{,\varphi
}^{2}-hh_{,\varphi \varphi }\right) \sin ^{2}\theta -h_{,\varphi }^{2}\cos
^{2}\theta }{\left( h^{2}\sin ^{2}\theta +h_{,\varphi }^{2}\right) ^{2}}.
\end{equation}

\begin{figure}[h]
\includegraphics[width=120mm,scale=0.7]{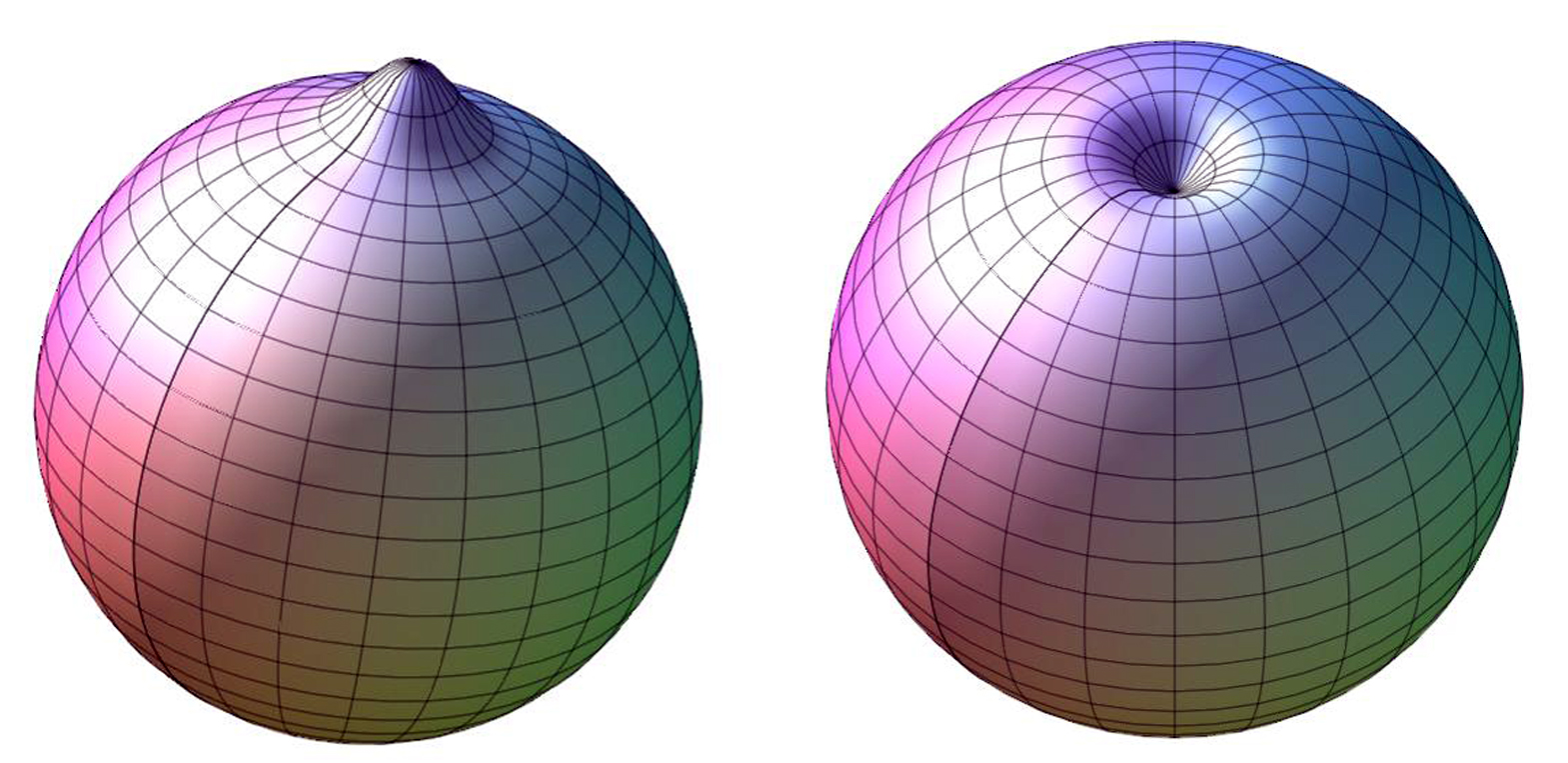}
\caption{A sphere with a bump in left and a miss / reverse bump on right.
The exact value of the parameters in Eq. (44) are as follow: $R=1,$ $\protect%
\alpha =40$ and $\protect\varepsilon _{0}=\pm 0.2.$ The positive and
negative $\protect\varepsilon _{0}$ give the left and the right
respectively. }
\label{fig: 1}
\end{figure}

The next special case is given by $h\left( \theta ,\varphi \right) =h\left(
\theta \right) $ with $h_{,\varphi }=0.$ The induced metric tensor becomes 
\begin{equation}
g_{ij}=\left[ 
\begin{array}{cc}
h^{2}+h_{,\theta }^{2} & 0 \\ 
0 & h^{2}\sin ^{2}\theta%
\end{array}%
\right]
\end{equation}%
with the second fundamental form%
\begin{equation}
K_{i}^{j}=\left[ 
\begin{array}{cc}
\frac{h^{2}-hh_{,\theta \theta }+2h_{,\theta }^{2}}{h^{3}\Delta ^{3/2}} & 0
\\ 
0 & \frac{\left( h^{2}+h_{,\theta }^{2}\right) \left( h\sin \theta
-h_{,\theta }\cos \theta \right) }{h^{4}\sin \theta \Delta ^{3/2}}%
\end{array}%
\right] .
\end{equation}%
Finally we find%
\begin{equation}
\kappa =\frac{\left( 3hh_{,\theta }^{2}+2h^{3}-h^{2}h_{,\theta \theta
}\right) \sin \theta -h_{,\theta }^{3}\cos \theta -h^{2}h_{,\theta }\cos
\theta }{2h\left( h^{2}+h_{,\theta }^{2}\right) ^{\frac{3}{2}}\sin \theta }
\end{equation}%
and%
\begin{equation}
\kappa _{G}=\frac{\left( h^{2}-hh_{,\theta \theta }+2h_{,\theta }^{2}\right)
\left( h\sin \theta -h_{,\theta }\cos \theta \right) }{h\left(
h^{2}+h_{,\theta }^{2}\right) ^{2}\sin \theta }
\end{equation}%
with%
\begin{equation}
\Delta =1+\frac{h_{,\theta }^{2}}{h^{2}}.
\end{equation}

\subsection{Example 1: A spheres with a bump}

\begin{figure}[h]
\includegraphics[width=120mm,scale=0.7]{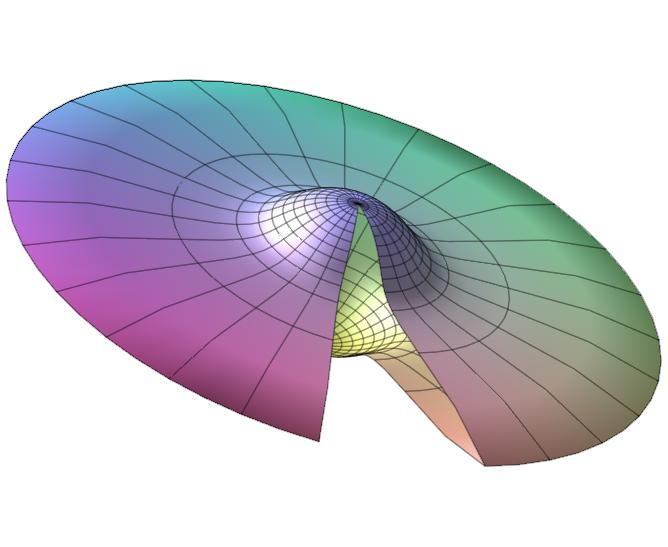}
\caption{A delta-dirac form of inflation at the equator of a sphere of
radius $R=1$. The other parameters in Eq. (50) are; $\protect\theta _{0}=%
\frac{\protect\pi }{2}$ and $\protect\zeta =0.1$. For a better visibility we
imposed $\protect\varphi \in \left[ 0,11\protect\pi /12\right] .$ }
\end{figure}


In Fig. 1 we plot a sphere with a bump, whose equation is given by%
\begin{equation}
h\left( \theta \right) =R\left( 1+\varepsilon _{0}\exp \left( -\alpha \theta
^{2}\right) \right)
\end{equation}%
in which $R$ is the radius of the background sphere and $\varepsilon _{0}$
and $\alpha $ are constants. Depending on the value of $\varepsilon _{0}$
and $\alpha $ the size of the bump changes and even get a reversed shape
with negative $\varepsilon _{0}$ (see Fig. 1 right). Using (41) and (42) we
find%
\begin{equation}
\kappa =\frac{\left( \lambda _{1}\varepsilon _{0}^{2}e^{-2\alpha \theta
^{2}}+\lambda _{2}\varepsilon _{0}e^{-\alpha \theta ^{2}}+\left( 1+\Delta
\right) \sin \theta \right) }{2R\left( 1+\varepsilon _{0}e^{-\alpha \theta
^{2}}\right) ^{3}\Delta ^{3/2}\sin \theta }
\end{equation}%
\begin{equation}
\kappa _{G}=\frac{\left( 1+\varepsilon _{0}e^{-\alpha \theta ^{2}}\left(
-4\alpha ^{2}\theta ^{2}+2\alpha +2\right) +\varepsilon _{0}^{2}e^{-2\alpha
\theta ^{2}}\left( 4\alpha ^{2}\theta ^{2}+2\alpha +1\right) \right) \left(
\varepsilon _{0}e^{-\alpha \theta ^{2}}\left( 2\alpha \theta \cos \theta
+\sin \theta \right) +\sin \theta \right) }{R^{2}\sin \theta \left(
1+\varepsilon _{0}e^{-\alpha \theta ^{2}}\right) \left( 1+2\varepsilon
_{0}e^{-\alpha \theta ^{2}}+\varepsilon _{0}^{2}e^{-2\alpha \theta
^{2}}\left( 4\alpha ^{2}\theta ^{2}+1\right) \right) ^{2}}
\end{equation}%
in which 
\begin{equation*}
\left( 
\begin{array}{r}
\lambda _{1} \\ 
\lambda _{2}%
\end{array}%
\right) =\left( 
\begin{array}{r}
\left( 2\alpha +4\alpha ^{2}\theta ^{2}+\Delta +1\right) \sin \theta
+2\Delta \alpha \theta \cos \theta \\ 
2\Delta \alpha \theta \cos \theta +\left( 2-4\alpha ^{2}\theta ^{2}+2\alpha
+2\Delta \right) \sin \theta%
\end{array}%
\right)
\end{equation*}%
and 
\begin{equation}
\Delta =1+\frac{4\varepsilon _{0}^{2}\alpha ^{2}\theta ^{2}e^{-2\alpha
\theta ^{2}}}{\left( 1+\varepsilon _{0}e^{-\alpha \theta ^{2}}\right) ^{2}}.
\end{equation}%
At $\theta =0$ one finds $\Delta =1$ and consequently%
\begin{equation}
\kappa =\frac{2\alpha \varepsilon _{0}+\varepsilon _{0}+1}{R\left(
1+\varepsilon _{0}\right) ^{2}}
\end{equation}%
and%
\begin{equation}
\kappa _{G}=\frac{\left( 2\varepsilon _{0}\alpha +\varepsilon _{0}+1\right)
^{2}}{R^{2}\left( 1+\varepsilon _{0}\right) ^{4}}=\kappa ^{2}.
\end{equation}

\subsection{Example 2: A sphere with a delta type inflation}

In our second example we consider 
\begin{equation}
h=R\left( 1+\frac{\zeta }{\left( \left( \theta -\theta _{0}\right)
^{2}+\zeta ^{2}\right) \pi }\right)
\end{equation}%
in which $\theta _{0},$ $\zeta $ and $R$ are constants. This model provides
a delta-type inflation at $\theta =\theta _{0}$ on a sphere of radius $R$
such that a smaller $\zeta $ produces more localized and sharper inflation.
In Fig. 2 we plot $h$ for $R=1,$ $\theta _{0}=\frac{\pi }{2}$ and $\zeta
=0.1 $. The sharp symmetric deformation at the equator of the sphere is
displayed in this figure.

\begin{figure}[h]
\includegraphics[width=120mm,scale=0.7]{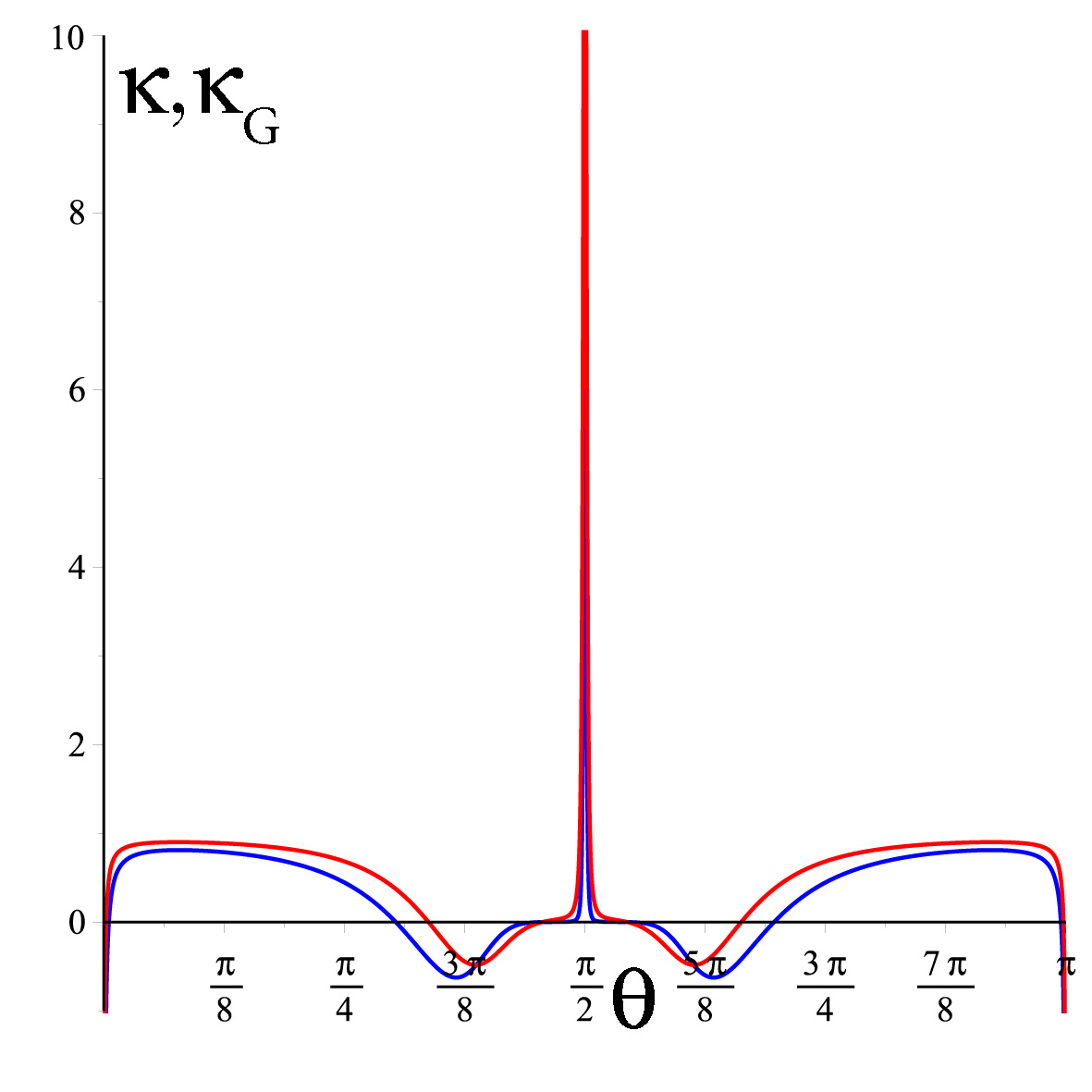}
\caption{$\protect\kappa $ (red/above) and $\protect\kappa _{G}$
(blue/below) in terms of $\protect\theta $ for the second example i.e., Eq.
(50). The specific values of parameters are as given in Fig. 2.}
\end{figure}


Using the general equations we find the total curvature given by%
\begin{equation}
\kappa =\frac{-x\zeta \left( \Lambda ^{2}\pi ^{2}H^{4}+4\zeta
^{2}x^{2}\right) \cot \theta +\pi H^{2}\Lambda \left( \Lambda ^{2}\pi
^{2}H^{4}-\pi \left( 3x^{2}-\zeta ^{2}\right) \zeta H\Lambda +6\zeta
^{2}x^{2}\right) }{R\Lambda \left( \Lambda ^{2}\pi ^{2}H^{4}+4\zeta
^{2}x^{2}\right) ^{\frac{3}{2}}}
\end{equation}%
in which $x=\theta -\theta _{0},$ $H=\left( \theta -\theta _{0}\right)
^{2}+\zeta ^{2}$ and $\Lambda =1+\frac{\zeta }{\pi H}.$ In addition, the
Gaussian curvature is also obtained as%
\begin{equation}
\kappa _{G}=\frac{\left( -6\Lambda \zeta \pi Hx^{2}+2\Lambda \zeta ^{3}\pi
H+\Lambda ^{2}\pi ^{2}H^{4}+8\zeta ^{2}x^{2}\right) \left( \Lambda \pi
H^{2}-2\zeta x\cot \theta \right) \pi H^{2}}{R^{2}\Lambda \left( \Lambda
^{2}\pi ^{2}H^{4}+4\zeta ^{2}x^{2}\right) }.
\end{equation}%
The limit of $\kappa $ and $\kappa _{G}$ when $\theta \rightarrow \theta
_{0} $ are found to be 
\begin{equation}
\lim_{\theta \rightarrow \theta _{0}}\kappa =\frac{1+\zeta ^{2}+\zeta ^{3}}{%
R\zeta \left( 1+\zeta \right) ^{2}},
\end{equation}%
\begin{equation}
\lim_{\theta \rightarrow \theta _{0}}\kappa _{G}=\frac{2+\zeta ^{2}+\zeta
^{3}}{R^{2}\left( 1+\zeta \right) ^{3}}
\end{equation}%
while their limits when $\zeta \rightarrow 0$ becomes $\frac{1}{R}$ and $%
\frac{1}{R^{2}}$, respectively, for all $\theta $ except for $\theta =\theta
_{0}$. At $\theta =\theta _{0}$ we find $\lim_{\zeta \rightarrow 0}\kappa
=\infty $ and $\lim_{\zeta \rightarrow 0}\kappa _{G}=\frac{2}{R^{2}}.$ In
Fig. 3 we plot $\kappa $ and $\kappa _{G}$ in terms of $\theta $ for the
specific choice of the parameters presented in Fig. 2.

\section{Conclusion}

Finding the effect of a small shape fluctuation on the free energy of a
membrane helps to predict the possible changes on the shape of such objects
due to diverse kind of external perturbations. For instance when one taps on
a soap bobble, it causes a change in its shape. We believe that the change
of the shape of the soap bobble follows the minimum change of its free
energy. Therefore knowing how to find the geometric properties of such
structure, makes our job simpler. Some of the $2-$dimensional surfaces can
be approximated as a local flat surface and any fluctuation is considered as
the MG. But in more general cases the unperturbed surface may not fit on a
flat surface neither locally nor globally. In such cases a different version
of the MG may be more helpful. In this work we have introduced the spherical
MG in its most general form. With some specific cases and examples we have
shown the application of this formalism. Our next step will be to find the
MG for toroidal and cylindrical surfaces.

\end{document}